\input psfig.sty

\documentclass{article} 
\begin{document}

\title{HST  as a powerful astrometric tool for pulsar astronomy.}

\author{R.P. Mignani$^1$ \and P.A. Caraveo$^2$ \and A. De Luca $^2$}

\date{ }

\maketitle

\noindent
$^1$ ST-ECF, Karl Schwarzschild  Str.2, D8574O   Garching   b.  Munchen

\noindent
$^2$ IFC-CNR,  Via Bassini  15,   I-20133 Milan

\vspace{10mm}

\begin{abstract}

Given their intrinsic faintness,  isolated neutron stars  represent an elusive  target  for optical  astronomy.   Up  to   date,  
an  optical counterpart has been identified  only for   a very  tiny fraction  of   the overall population of  more than 1000 
radio  pulsars. The HST contributions to the optical studies  of isolated neutron    stars  have been  recently reviewed   
by Mignani et al. (2000a)\cite{13} and by Caraveo et al. (these proceedings). Here, we review its specific contribution in the 
field of pulsar astrometry and kinematics.
\end{abstract}

\section{Introduction.}
It is generally accepted that isolated pulsars are born during Type-II supernova explosions following the core collapse 
of the massive progenitor. If the explosion is asymmetric, the collapsed core of the progenitor, i.e. the newborn neutron 
star,  is expected to acquire a recoil velocity. Isolated pulsars  are thus characterized by an intrinsic  proper motion on 
the plane of the sky, which scales directly with the ratio between the pulsar projected velocity and its distance.

For this reason,  local pulsars have been the first natural targets for proper motion measurements in the radio band. In 
most cases such measurements are pursued with the aim of confirming the putative association of a pulsar  with a 
supernova remnant observed along the line of sight. On a wider scale, kinematic studies are pursued to derive  neutron 
star velocities and thus to obtain useful information on the dynamics of the SN explosion. So far, radio proper motions 
have been measured for about 100 pulsars, and the derived velocity distribution peaks around 400 km/s (Lyne \& 
Lorimer 1994)\cite{9}.

With the detection of neutron stars in the optical domain, kinematic studies of pulsars became possible using classical 
optical astrometry technics. This approach was successfully applied to those pulsars for which no radio proper motion 
measurements were available (the Crab pulsar) or possible (the radio-silent pulsar Geminga),  or for which they 
produced discrepant (the Vela pulsars) or highly uncertain (PSR0656+14) results. In some cases (Geminga and 
PSR0656+14), optical proper motion was used as a tool to confirm the proposed identification,  while waiting for a 
reliable optical timing.

After the seminal work carried out from the ground  with ESO telescopes (see Mignani et al. 2000b\cite{14} for a review), HST 
observations have made it possible a major step forward  in the optical astrometry of pulsars. In particular,  the excellent 
angular resolution of the WFPC2 together with its acceptably large field of view, allowed to perform very precise 
astrometric studies, yielding proper motion measurements for nearly half of the pulsars which have an optical 
counterpart. While in most cases WFPC2 observations represented a follow-up of work already carried out from the  
ground  they provided  measurements of equal, or even higher, accuracy in a much shorter time span, which made them 
easily repeatable and thus reliable (see De Luca et al. these proceedings). The proper motion of pulsars measured by the 
HST are summarized in Table 1, together with values obtained earlier on either in the radio or in the optical domains. 
In the following sections, the major achievements obtained by the HST in the field of pulsar astrometry are reviewed. A 
comprehnsive description of  the details of WFPC2 astrometry and of the approach to the data analysis is given in the 
companion paper by De Luca et al (these proceedings).

\begin{table}
\small
\begin{tabular}{cccccccc} \hline \hline \\
Pulsar & mag & $\mu_{\alpha}$(mas/yr) & $\mu_{\delta}$(mas/yr) & P.A. & Time span & Origin & References \\ \hline
Geminga & 25.5 & $149\pm44$ & $109\pm44$ & $54\pm13$ & 10 yrs & Opt-ground & Mignani et al. (1994)\cite{12} \\
 & & $138\pm4$ & $97\pm4$ & $55\pm2$ & 1 yr & HST & Caraveo et al. (1996)\cite{4} \\ \hline
Crab & 16.6 & $-13\pm2$ & $7\pm3$ & $298\pm10$ & 77 yrs & Opt-ground & Wyckoff\&Murray (1977)\cite{23} \\
 & & $-17\pm3$ & $7\pm3$ & $292\pm10$ & 1.7 yrs & HST & Caraveo\&Mignani (1999)\cite{6} \\ \hline
Vela & 23.6 & $-40\pm4$ & $28\pm2$ & $304\pm3$ & 2 yrs & Radio & Bailes et al. (1989)\cite{1} \\
 & & $-46\pm2$ & $24\pm2$ & $297\pm2$ & 2 yrs & HST &De Luca et al. (2000)\cite{7} \\ \hline
PSR0656+14 & 25 & $73\pm20$ & $-26\pm13$ & $109\pm10$ & 4 yrs & Radio & Pavlov et al. (1996)\cite{17} \\
 & & $43\pm2$ & $-2\pm3$ & $93\pm4$ & 4 yrs & HST & Mignani et al. (2000c)\cite{15} \\ \hline \hline
\end{tabular}

\caption{Summary of the pulsars' proper motions measured by the HST. The first column lists the pulsar name and the 
second its V-band magnitude.The pulsars are sorted according to the epoch of HST observations. Proper motions values 
along right ascension and declinations are reported in columns three and four, respectively in units of milliarcseconds 
per year. The proper motion position angle, in degrees, is in column five. The time span covered by the observations 
used for the proper motion measurements is given in column six. The numbers in bold correspond to HST 
measurements, while the others are the reference values obtained either in  radio or in the optical with ground-based  
telescopes (column seven). For each measurement, the references are quoted in the last column.} 
\end{table}
\normalsize

\section{Geminga}
Geminga is the first neutron star for which a WFPC2 proper motion was obtained (Caraveo et al. 1996\cite{4}). The proper 
motion of Geminga was originally measured from the ground (Bignami et al. 1993\cite{2}) and was used to strengthen  its 
proposed optical identification. The measurement was later refined by Mignani et al. (1994)\cite{12} using a 10 years timeline of 
ground based observations. Both results were significantly improved  by the proper motion measurement obtained by 
the WFPC2 ($\mu_{\alpha} =138 \pm4$ mas/yr and $\mu_{\delta} =97 \pm4$). The reliability of this value has been recently reassessed by DeLuca 
et al (these proceedings) using two independent couples of WFPC2 observations.
 
Together with the measurement of the pulsar proper motion, WFPC2 observations were used by Caraveo et al. (1996)\cite{4} to 
compute its  parallactic displacement ($6 \pm2$ mas), the only one ever obtained in the optical domain for a neutron star. 
The pulsar parallax thus provided  for the first time the  value of the neutron star distance (157 pc), in perfect agreement 
with the constraints derived from X-ray observations.

The same data were also used by Caraveo et al. (1998)\cite{5} in a multi-step astrometric chain aimed at linking the very 
accurate WFPC2 relative position of Geminga to the Tycho/Hipparcos absolute reference frame. This procedure yielded 
the Geminga's absolute coordinates with a precision of 40 mas i.e. with an accuracy higher that then one achievable for 
most radio pulsars including the Crab. The precise absolute position of Geminga, coupled with the accurate proper 
motion measurement, made it possible to phase together gamma-ray observations obtained years apart  and to perform a 
more accurate timing analysis of the pulsar (Mattox et al. 1998\cite{11}).

\section{The Crab Pulsar.}
A further  demonstration of the outstanding capabilities of the WFPC2 for very accurate astrometry measurements came 
with the new assessment of the proper motion of the Crab pulsar (Caraveo \& Mignani 1999\cite{6}). So far, the proper motion 
of the Crab Pulsar has been obtained only in the optical. After the original work performed by Trimble (1968)\cite{22}, the Crab 
proper motion was reliably measured by Wyckoff \& Murray (1977)\cite{23} using photographic plates covering a time span of 
77 year. Their value (see Tab.1) has been the reference ever since.

The availability of high resolution ROSAT images, showing X-ray jest apparently aligned with the proper motion 
direction, prompted Caraveo \& Mignani (1999)\cite{6} to obtain a new independent measurement of the Crab pulsar proper 
motion. Using a set of archival WFPC2 observations, collected over a period of 1.7 years, Caraveo \& Mignani (1999)\cite{6} 
obtained a value of $\mu_{\alpha} = -17\pm3$ mas/yr and  $\mu_{\delta}= 7\pm3$ mas/yr, i.e. fully consistent with that of  Wyckoff \& Murray 
(1977)\cite{23} and of comparable accuracy, in spite of the much shorter time span of less than two years. The proper motion 
direction, now confirmed by two independent measurements, turns out to be intriguingly similar to the  axis of 
simmetry  of  the torus observed in X-rays by ROSAT (Hester et al. 1996)\cite{8}. This coincidence, further stressed by recent 
Chandra high resolution images (Fig.1), led Caraveo \& Mignani (1999)\cite{6} to speculate about possible connections 
between the pulsar dynamics and the observed morphology of the inner Crab Nebula.

\begin{figure}
\centerline{\hbox{\psfig{figure=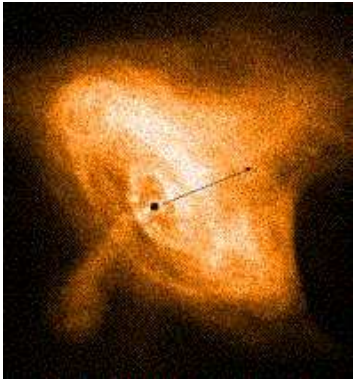,height=5cm,clip=}}}
\caption{X-ray image around  the Crab pulsar taken with the High Resolution 
Camera of the Chandra X-ray Observatory. Both the torus and thejet/counter-jet
 are now clearly resolved. The arrow indicates the direction ofthe pulsar 
proper motion measured by Caraveo \& Mignani (1999)[6]} 
\end{figure}

\section{The Vela Pulsar.}
The measurement of the Vela pulsar proper motion has been performed several times, both in radio and in the optical, 
using different instruments and technics. However, while proper motion measurements performed in the optical turned 
out to be consistent, the ones performed in radio  yielded conflicting results (see Nasuti. et al. 1997\cite{16} for a summary of 
the available measurements). An independent assessment of the Vela pulsar proper motion  through WFPC2 astrometry 
was needed. Over the years, repeated  WFPC2 observations of the Vela field were obtained as a part of a larger program 
aimed at the assessment of the pulsar's optical parallax (see De Luca et al-these proceedings). These yielded an accurate 
determination of the Vela pulsar proper motion of $\mu_{\alpha} =-46 \pm2$ mas/yr and $\mu_{\delta}=24 +/2$ mas/yr (De Luca et al. 2000\cite{7}), a result 
slightly better than the previous ground based one obtained by Nasuti et al. (1997)\cite{16} using observations covering 20 
years.

As in the case of the Crab pulsar, the proper motion can be used to investigate pulsar/SNR interactions. In particular, the 
direction of the proper motion appears to be aligned with the axis of symmetry of the compact X-ray nebula observed 
around the pulsar (Markwardt \& Ogelman 1998\cite{10}). The nebula has been  now resolved by high resolution Chandra 
images  to consist of  a toroidal structure crossed by a jet-like feature apparently originating from the pulsar (Pavlov et 
al. 2000\cite{19}).  In the case of the Crab, both the axis of symmetry of the torus and the jet are aligned with the direction of the 
pulsar proper motion (see Fig.2). Indeed, while for the Crab it is difficult to quantify the precision of the alignment, 
owing to the complex structure of the X-ray jets, in the case of Vela the alignment between the X-ray jets and the pulsar 
proper motion can be assessed precisely. Pavlov et al. (2000)\cite{19} quote an uncertainty of 1-2 degrees (i.e. as good as it can 
be), pointing to a clear correlation between the kinematics of the pulsar and its surrounding structures.

\begin{figure}
\centerline{\hbox{\psfig{figure=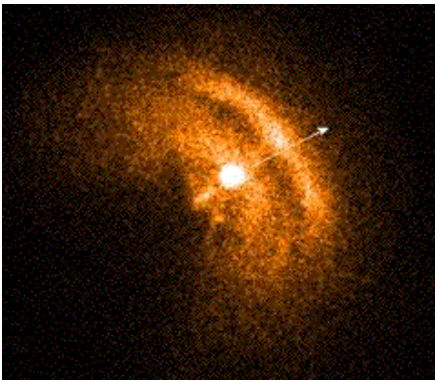,height=5cm,clip=}}}
\caption{X-ray image around  the Vela pulsar taken with the High Resolution 
Camera of the Chandra X-ray Observatory  (Pavlov et al. 2000[19]). For the 
first time, the compact X-ray nebula around the pulsar has been resolved into 
a  torus 
 apparently crossed by a jet-like structure. The arrow indicates 
the direction of the pulsar proper motion measured by De Luca (2000)[7]. The 
similarity with the Crab (Fig.1) is evident.}
\end{figure}

\section{PSR0656+14.}
The most recent pulsar astrometry result obtained by the HST is the measurement of the proper motion of  the candidate 
counterpart to PSR0656+14  (Mignani et al. 2000c\cite{15}). A radio proper motion measurement of the pulsar was originally 
performed from the NRAO Very Large Array by Thompson and Cordova (1994)\cite{21} to investigate its possible association  
with the nearby supernova remnant known as Monogem Ring. A proper motion value of $\mu_{\alpha} =  64 \pm 11$ mas/yr and $\mu_{delta} =-28 \pm 4$ mas/yr was obtained. Pavlov, Stringfellow and Cordova (1996)\cite{17} reevaluated the systematic errors affecting 
the measurements and revised the proper motion  in $\mu_{\alpha} =73 \pm20$ mas/yr and $\mu_{\delta} =-26 \pm13$ mas/yr (PA= $109 \pm 10$ 
deg), a value hinting the presence of a proper motion but too uncertain to be conclusive.

Using two sets of WFPC2 observations taken 4 years apart, Mignani et al. (2000c)\cite{15} measured a relative displacement 
corresponding to a proper motion of $43 \pm 3$ mas/yr, virtually all along right ascension (PA of 93 degrees). Since the 
two observations were taken at the same period of the year,  the measurement of the pulsar displacement is not affected 
by any possible parallactic displacement. Although the WFPC2 result does not represent the first proper motion 
measurement of PSR0656+14,  it is by far the most reliable.

For the allowed range of the pulsar radio distance (300-800 pc) the measured proper motion would imply a transverse 
velocity between 50 and 100 km/s, a value which, although rather low, is typical of neutron stars. Such velocity, 
coupled with the unusual colors of the object (Pavlov et al. 1997\cite{18}), virtually identifies it as an isolated  neutron star. 
Independently of the optical timing (Shearer et al. 1997\cite{20}), which is still to be confirmed, the proper motion measurement  
provides a new and robust confirmation of the proposed pulsar identification.

\section{Conclusions.}
So far, HST allowed to obtain proper motion measurements for 4  pulsars. Irrespective of the time span covered by the 
observations, HST yielded the most accurate proper motion measurements ever. Moreover,  the accuracy achieved  
(Tab.1) is poorly dependent on the object brightness. Indeed, similar values have been obtained for the Crab and for the 
10 magnitudes fainter Geminga.

The astrophysical information which can be obtained from proper motion studies involve different aspects of pulsar 
astronomy. First, as was the case for Geminga and more recently for  PSR0656+14 a proper motion is a powerful tool 
either to confirm the optical identification of a pulsar already known to move or to claim a new identification on the 
basis of the fast moving nature of the candidate.

Second, as shown in the cases of the Crab and Vela pulsars, kinematic studies give a new handle to investigate dynamic 
interactions between the pulsar and the inner regions of the surrounding SNRs and may be to clarify the mechanism(s) 
responsible for the neutron star kick.

Last but not least, optical proper motions of pulsars with known distance, yield  tranverse velocity measurements to be 
compared with the radio ones. The values of the transverse velocities computed for all the pulsars with an HST proper 
motion are summarized in Tab.2 for the nominal values of the  objects' distances. It is easy to note that all the nearby 
pulsars (Vela,  PSR0656+14 and Geminga) are moving at a speed significantly smaller than the average value (400 
km/s) obtained for a much larger sample of radio pulsars. Of course, many more neutron stars must be observed in the 
optical in order to establish whether this is just a mere coincidence or it is due to an unknown observational bias.

\begin{table}
\begin{tabular}{cccc} \hline \hline \\
Pulsar & PM  (mas/yr) & v (km/s) & d (pc) \\ \hline
Vela & $52\pm3$ & 130 & 500 \\
Geminga & $170\pm6$ & 133 & 157 \\
PSR0656+14 & $43\pm2$ & 163 & 760 \\
Crab & $18\pm3$ & 180 & 2000 \\ \hline \hline
\end{tabular}
\caption{Tranverse velocity inferred for pulsars with a  proper motions measured by the HST. The first column gives the 
pulsar name, the second the total proper motion in milliarcseconds per year, the third the inferred transverse velocity 
for the nominal pulsar distance (column four).The pulsars are sorted according to increasing value of the velocity.}
\end{table}

\end{document}